# Pricing decisions under manufacturer's component open-supply strategy

Peiya Zhu[1], Xiaofei Qian[1, 2, 3, *], Xinbao Liu[1, 2, 3, *], Shaojun Lu[1, 2, 3]

[1] School of Management, Hefei University of Technology, Hefei 230009, PR. China

[2] Key Laboratory of Process Optimization and Intelligent Decision-making of the Ministry of Education, Hefei 230009, PR. China

[3] Ministry of Education Engineering Research Center for Intelligent Decision-Making and Information System Technologies, Hefei 230009, PR. China

**Abstract:** Faced with huge market potential and increasing competition in emerging industries, product manufacturers with key technologies tend to consider whether to implement a component open-supply strategy. This study focuses on a pricing game induced by the component open-supply strategy between a vertically integrated manufacturer (who produces key components and end products) and an exterior product manufacturer (who produces end products using purchased key components) with different customer perceived value and different cost structure. This study first establishes a three-stage pricing game model and proposes demand functions by incorporating relative customer perceived value. Based on the demand functions, we obtain feasible regions of the exterior manufacturer's sourcing decision and the optimal price decision in each region. Then the effects of relative customer perceived value, cost structure, and market structure on price decisions and optimal profits of the vertically integrated manufacturer are demonstrated. Finally, as for the optimal component supply strategy, we present a generalized closed-supply Pareto zone and establish supply strategy Pareto zones under several specific configurations.

[*] Corresponding authors.

E-mail address: peiyazhu@126.com (P. Zhu), qianxiaofei888@126.com (X. Qian), lxb@hfut.edu.cn (X. Liu), lushaojun@hfut.edu.cn (S. Lu).

# 1. Introduction

An emerging industry is usually formed by the industrialization of new key technologies, and the key technology owner can play one of the following three roles in the market (Venkatesh, Chintagunta, and Mahajan 2006; Xu, Gurnani, and Desiraju 2010): (1) A component manufacturer who only produces and supplies key components and does not enter the terminal market directly, e.g., Panasonic is one of the largest suppliers of electric vehicle batteries worldwide who does not sell electric vehicles. (2) A product manufacturer who builds and markets its own branded end products using the key components produced by itself or other component manufacturers, e.g., BYD Auto used to be an electric vehicle manufacturer who possesses the ability to develop and produce almost all of the key components including power batteries. (3) A "dual manufacturer" who sells branded end products while providing key components to other product manufacturers, e.g., Samsung promotes its own branded phone and supplies key components such as screens and processors for smartphone manufacturers.

Key component markets are profitable and demanding more technological innovations with the significant expansion of emerging industrial markets. An example is the new energy vehicle industry. Electric car sales have soared in the past decade under the consistent objectives of environmental friendliness and sustainability. Sales of new electric vehicles worldwide topped 2.1 million in 2019 and boost the stock to 7.2 million. It is projected that the global electric vehicle battery capacity increases from around 170 GWh per year today to 1.5 TWh per year in 2030 (Global Electric Vehicle Outlook, 2020). The ongoing demanding trend of low cost and high-performance power batteries is predicted to continue in the future, thus, more and more enterprises pay attention to the bright prospect of the battery value chain and launch a series of the layout.

To cope with the increasingly fierce market competition and keep up with the rapid development of emerging industries, manufacturers with key component technology should change their role flexibly according to internal and external circumstances. Take an example of China's biggest electric vehicle manufacturer Build Your Dreams (BYD Auto). According to the China Automotive Industry Association, the national top ten manufacturers of new energy vehicle in 2017 except BYD Auto need to buy batteries from the third-party battery suppliers. The same year, BYD Auto announced its plan of changing from "vertical integration" to "reform and opening". That is, BYD Auto will split its power battery business from a previously closed self-sufficiency strategy to an open-supply strategy and join in the power battery market. Therefore, to improve its competitiveness in the battery market, BYD Auto needs to invest in building high-capacity production lines of power batteries and keeping the battery technology advanced. By implementing the power battery open-supply strategy, BYD Auto may transform its role from a product manufacturer to a "dual manufacturer" or even a component manufacturer in some special conditions.

When a product manufacturer with key technologies enlarges production capacity and opens the component supply, the manufacturer will face intensified end product market competition with other product manufacturers who equip their products with the key components of the same brand and quality. For example, the performance of power battery (e.g., endurance mileage, cycle life, charging rate, etc.) is the main concerns of customers when purchasing electric vehicles, we believe that different brands of electric vehicles equipped with the same brand and performance level of power batteries will be classified into the same market segment. Thus, in terms of whether to take the open-supply strategy, the product manufacturer needs to balance the increased revenue from component market expansion against the decreased revenue due to end product market share loss through the pricing decisions of key component and product. Meanwhile, the product manufacturer who relies on component outsourcing should also weigh the pros and cons of changing the supplier partner and determine product prices promptly.

The above discussion outlines the competition between two manufacturers which is induced by the open-supply strategy. In reality, many important factors will influence the manufacturer's strategy decision, such as the original market share, the market spillover after the implementation of the strategy, customer perceived value of products, and the manufacturing cost of the same industry. These factors further affect the pricing behavior of manufacturers. In this paper, we model a three-stage pricing game between a vertically integrated manufacturer (who produces key components and end products) and an exterior product manufacturer (who produces end products using purchased key components) by considering the above factors. Based on the model, we aim to solve the following two problems in this paper:

(1) When should a vertically integrated manufacturer open the component supply? How to develop price policies?

(2) How does the exterior product manufacturer react faced to the component open-supply strategy?

To the best of our knowledge, we are the first one to study the two manufacturers' competition induced by component open-supply strategy under different customer perceived value, market spillover, and different cost structures. Moreover, we obtain several meaningful insights that can provide useful decision supports for the product manufacturers who own key technologies and exterior product manufacturers. For comparison, we discuss two scenarios: the vertically integrated manufacturer has a higher customer perceived value (i.e. $0 < \theta < 1$) or a lower customer perceived value (i.e. $\theta > 1$) than that of the exterior manufacturer. First, we derive the optimal purchasing strategy and the optimal sale price of the exterior manufacturer. Second, we analyze the factors that influence the feasible decision region of the vertically integrated manufacturer when the exterior manufacturer enters the market. The larger the common market size, the more likely an exterior manufacturer will be attracted in both the two scenarios. Besides, in scenario $\theta > 1$, the decrease of the exterior manufacturer's operation cost will also increase the likelihood. Third, we yield the optimal prices and profits of the vertically integrated manufacturer. In scenario $0 < \theta < 1$, the equilibrium price decisions of the vertically integrated manufacturer are influenced by the common market size and the operation cost of two manufacturers. However, in scenario $\theta > 1$, the equilibrium price decisions of the vertically integrated manufacturer are not affected by its own operation cost. Finally, a generalized closed-supply Pareto zone and supply strategy Pareto zones under several sets of specific configurations are presented. We examine how the change of the relative customer perceived value, the operation cost of each manufacturer, the original market of the vertically integrated manufacturer, and the market spillover influence the supply strategy Pareto zones.

The rest of the paper is organized as follows. We position our study in Section 2 and introduce the model setting in Section 3. The analysis and main results of the model are presented in Section 4. We conclude the paper and indicate the potential future research directions in Section 5. For the sake of clarity, all proofs are provided in Appendix B.

## 2. Literature review

For the theoretical basis and model conceptualization, we draw on the following research streams: pricing strategy in the multichannel distribution supply chain, market segmentation, customer choice and ingredient branding.

### 2.1. Pricing strategy in multichannel distribution supply chain

In the past few years, the multichannel distribution has gained much attention from the supply chain management research. The pricing strategy of the multichannel distribution in the supply chain has also been studied quite extensively. These literature focuses on channel structure choice(Balasubramanian 1998; Chiang, Chhajed, and Hess 2003; Tsay and Agrawal 2004; Yao and Liu 2005; Cattani et al. 2006; Liu, Gupta, and Zhang 2006; Dumrongsiri et al. 2008; Matsui 2016; Matsui 2017; Matsui 2020; Guo et al. 2020;

Zhang, Yao, and Xu 2020), service competition (Dan, Xu, and Liu 2012; Li and Li 2016; Dan, Zhang, and Zhou 2017; Liu et al. 2019; Pun, Chen, and Li 2020) and dual-channel closed-loop supply chain (Ma, Zhao, and Ke 2013; Saha, Sarmah, and Moon 2016; Gan, Pujawan, and Widodo 2017). A number of the published papers on this subject address the scenario of a single product sold in three possible supply chain structures, namely, traditional brick-and-mortar retail channel only, the direct channel through the Internet, or dual channel. Balasubramanian (1998) presents a set of insights on how the entry, market structure, and customer information influence the competition between direct marketers and conventional retailers. Chiang, Chhajed, and Hess (2003) find that adding a direct marketing channel indirectly helps both the retail and manufacturer improve profitability. Matsui (2016) shows that even the symmetric manufacturers choose an asymmetric distribution policy in the dual-channel supply chain. Further, Matsui (2017) and Matsui (2020) investigates the timing problem of when should a manufacturer post its direct price and bargain a wholesale price with a retailer in dual-channel supply chains, which suggests that the manufacturer should reveal the direct price before or upon setting the wholesale price. Most of the literature in this stream analyzes the economic impact of the introduction of the direct online channel. This study discusses the impact of the open-supply strategy on two heterogeneous manufacturers, which is similar to the model of the introduction of an indirect channel. Moreover, in contrast to many studies that analyze the multichannel distribution of a single product in the supply chain, we model a dual-channel of two different branded products that compete for a common customer group, and only the vertical integrated manufacturer chooses the distribution channel structure.

*2.2. Market segmentation and customer choice*

The literature on market segmentation typically considers that customers are heterogeneous in their valuation of the product quality (Moorthy 1984). Further, Lambertini (2018) propose a dynamic model where two vertically related firms invest in development efforts for quality improvement, and customers are characterized by a level of marginal willingness to pay for quality. Debo, Toktay, and Van Wassenhove (2005) investigate the joint pricing and production technology selection strategy of a manufacturer who attempts to bring remanufacturing products into a market with heterogeneous consumers. They propose a class of distribution of the market form $F(\theta) = 1 - (1-\theta)^k$, where $k \in (0,\infty)$, denoting the number of consumers with a willingness to pay in $[0,\theta]$. The uniform distribution is a special case of this distribution when setting $k = 1$. Kumar and Ruan (2006) consider that heterogeneous customers are segmented into brand loyal and store loyal. Luo et al. (2017) model a market that individual customer perceives the good brand better than the average brand. Ferrer and Swaminathan (2006) and Ma, Gong, and Mirchandani (2020) consider that consumers have lower quality valuations for remanufactured products. Some researchers also investigate the customer choice in dual-channel distribution supply chain contexts (Reardon and McCorkle 2002; Chen, Kaya, and Özer 2008). We adopt the frequently used assumption that the customer is heterogeneous and uniformly distributed in their perceived value of end products.

*2.3. Ingredient branding*

Researches on ingredient branding take end product manufacturers' perspective or ingredient suppliers' perspective on creating a brand for a component for higher quality or performance of the component and the end product (Venkatesh and Mahajan 1997). Lienland, Baumgartner, and Knubben (2013) study the significance of a manufacturer's reputational change by ingredient branding of complex products with high manufacturing intensity. Zhang et al. (2013) investigate ingredient branding strategies where the supplier and the original equipment manufacturer form a brand alliance. Except for the studies on ingredient branding of tangible products, Helm and Özergin (2015) focus on the ingredient service branding. They show an additional value added to the buyers' perception of the end-product service quality at the presence of an ingredient service brand even if the host brand is of low quality. This stream of the literature suggests that enterprises should focus on developing core competencies rather than vertical integration. In this paper, we employ the concept of ingredient branding in our component open-supply strategy. We model that a vertically integrated manufacturer considers creating a brand for the key component and opening supply to enter the component market.

Venkatesh, Chintagunta, and Mahajan (2006) and Xu, Gurnani, and Desiraju (2010) are also related to our study. The former considers how the degree of heterogeneity in consumer preference and brand-specific strengths influence the supply chain structure choice and pricing strategy of a proprietary component manufacturer. Further, the latter extends the model by considering a capability advantage on the original equipment manufacturer in bringing the end product to market. These are somewhat different from this study. At first, their analyses are based on the assumption that the proprietary component manufacturer is the sole source of component supply while we assume the component market is not monopolistic. Thus we need to consider the procurement decision of the exterior manufacturer. Second, they develop a spatial competition model between two end-product brands and consider the superiority or inferiority of the location-specific brand value, while we ignore the spatial factor and focus on the relative customer preference of two branded products. At last, the former treats the cost structure of two competing manufacturers as the same and the latter considers a cost advantage of one manufacturer, while we generalize the cost structures of two manufacturers and analyze how they impact the decisions of both manufacturers.

In sum, while our study is motivated by the mentioned literature streams, it is different from and arguably complementary to extant perspectives.

## 3. Model setting

In this section, we present the model assumptions and game sequence in SubSection 3.1 and derive the demand functions in SubSection 3.2. All notations are listed in Table 1.

Table 1. Notations

| Notations | Definitions |
|---|---|
| $w$ | The unit wholesale price of the proprietary component of supplier $S$ |
| $p_i$ | The unit sale price of $M_i$'s end product in the open-supply context |
| $p_e$ | The unit sale price of $M_e$'s end product in the open-supply context |
| $c$ | The unit production cost of the proprietary component of the supplier $S$ |
| $m_i$ | The unit assembly and sale cost of $M_i$'s end product |
| $m_e$ | The unit assembly and sale cost of $M_e$'s end product |
| $Q_s$ | The component demand of supplier $S$ |
| $Q_i$ | The end product demand of $M_i$ |
| $Q_e$ | The end product demand of $M_e$ |
| $v$ | The customer perceived value of $M_i$'s end product |
| $\theta$ | The relative customer perceived value of $M_e$'s end product compared to $M_i$'s end product |
| $A$ | The original market share of $M_i$'s end product |
| $\hat{A}$ | The common end product market share of two manufacturers who use the component from $S$ in the open-supply context |
| $\gamma_1$ | The market spillover degree after the implementation of component open-supply strategy |
| $\gamma_2$ | The proportion of the exterior market share that $M_e$ occupies in open-supply context |
| $p_i^0$ | The unit sale price of $M_i$'s end product in closed-supply context |
| $p_e^0$ | The unit sale price of $M_e$'s end product equipped with a core component of other supplier in |

| | open-supply context |
|---|---|
| $w_0$ | The unit purchase price of the other supplier's component |
| $\Pi_{M_i}^0$ | The profit of $M_i$ in original closed-supply context |
| $\pi_{M_e}^0$ | The profit of $M_e$ equipped with a core component of other supplier in open-supply context |
| $\Pi_{M_i+s,r}$ | The total profit of $S$ and $M_i$ in open-supply context in stage $r$, $r=\{1,2,3\}$ |
| $\pi_{M_e,r}$ | The profit of $M_e$ equipped the component of supplier $S$ in stage $r$, $r=\{1,2,3\}$ |

## 3.1. Assumptions and game sequence

**Industry structure**. We model a vertically integrated manufacturer (e.g., BYD Auto) that comprises a proprietary component supply department (take as a supplier $S$ afterwards) and an end-product manufacturing department (take as an interior manufacturer $M_i$ afterwards) who makes its own sales to the market. The end product (e.g., an electric car) is made of a proprietary component (e.g., an electric battery) and an installation base (e.g., an electric car assembly system). The proprietary component is a cutting-edge innovation but the vertically integrated manufacture is not the sole supply source. To seize the component markets and pursue higher profits, the vertically integrated manufacturer considers implementing the component open-supply strategy. When the component supply opens, the other manufacturer (take as an exterior manufacturer $M_e$ afterwards) who turns to order component from $S$ will compete for the alternative end product with $M_i$ in the downstream market. Assume that both manufacturers who use the component from $S$ produce the same kind of product. Using a component from a high-reputation supplier for product innovation improves the perception of a final offering (Linder and Seidenstricker 2018). We assume that the end product market is divided into distinct segments by core component technology, so the two branded end products target the same market after $M_e$ turning to $S$. The investment of building component brand is $K$. $S$ incurs the production cost of $c$ per unit, and we normalize $c=0$ without loss of generality. $M_i$ and $M_e$ cost $m_i$ and $m_e$ per unit to assemble and sell the end product respectively.

**Market structure**. The size of the end product market is normalized to one. Initially, the market share of the component and end product among different brands are formed. We model the original market share of $M_i$ with $A$, and the original market share of other manufacturers (except $M_i$) is $1-A$. Once the vertically integrated manufacturer opens the component supply, its end product market share expands to $\hat{A}=A+\gamma_1(1-A)$ due to the improved brand reputation of key components, where $\gamma_1\in[0,1)$ denotes the market spillover degree. If no exterior manufacturer turns to $S$, $M_i$ monopolize the market segment $\hat{A}$. If $M_e$ who occupies $\gamma_2(1-A)$ market share after the market spillover is attracted to $S$, the market size of $S$ remains $\hat{A}$ and the two manufacturers compete in common market $\hat{A}$ by pricing strategy. $\gamma_2$ denotes the proportion of the exterior market share that $M_e$ occupies in open-supply context. The parameters satisfy $\gamma_2 \in (0,1)$ and $\gamma_1 + \gamma_2 \leq 1$.

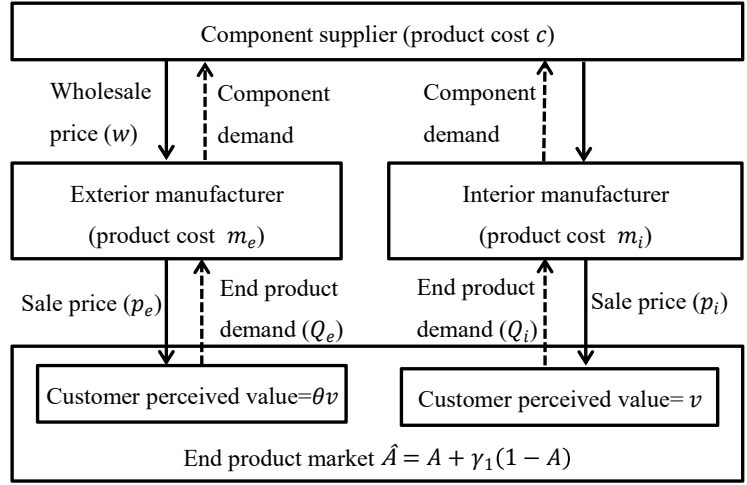

Fig. 1. Competition between manufacturers in open-supply context

**Customer Preferences.** Consumers typically differ in their willingness to pay. For this reason, we associate with each consumer's perceived value for a product of a specific brand. As frequently used in the market segmentation literature (Chiang, Chhajed, and Hess 2003; Ma, Zhao, and Ke 2013), we assume that the customer perceived value $v$ follows the uniform distribution within the customer population on $[0, 1]$. For each customer, the perceived value of brand $M_i$ is $v$, and the perceived value of brand $M_e$ is $\theta v$. The parameter $\theta$ is used to measure the relative customer perceived value of $M_e$'s product compared to $M_i$'s product. We allow $\theta$ to be less or greater than 1 in terms of the discrepant brand reputation, service quality, or marketing capability between $M_i$ and $M_e$. We assume that the customer perceived value of two differentiated products is not identical, i.e. $\theta \neq 1$. A customer only purchases a product if the perceived value on that product is greater than or equal to the product price. Fig. 1 illustrates the competition between two manufacturers in open-supply context.

The optimal supply/sourcing strategy and pricing problem are assumed to be a Stackelberg pricing game. Both agents are risk-neutral and aim to maximize their expected profits. The sequence of events is summarized as follows and illustrated in Fig. 2:

Stage 1 (Vertically integrated manufacturer's supply strategy). The game starts with the vertically integrated manufacturer as the leader, announcing whether to open its component supply. If the supply is open, then $M_e$ can adopt the component of $S$ and compete with $M_i$ in the common downstream market. Otherwise, the respective original market share of the two end products remains stable.

Stage 2 (Vertically integrated manufacturer's price decision). Once the vertically integrated manufacturer opens its component supply, it needs to set the wholesale price of components and adjust the sale price of the end product simultaneously and collusively after predicting the behavior of $M_e$.

Stage 3 (Exterior manufacturer's sourcing strategy and price decision). By considering the component wholesale price and product sale price of the vertically integrated manufacturer, $M_e$ chooses its sourcing strategy and sale price. Finally, the demand of the two manufacturers' end products and the component of supplier $S$ are realized.

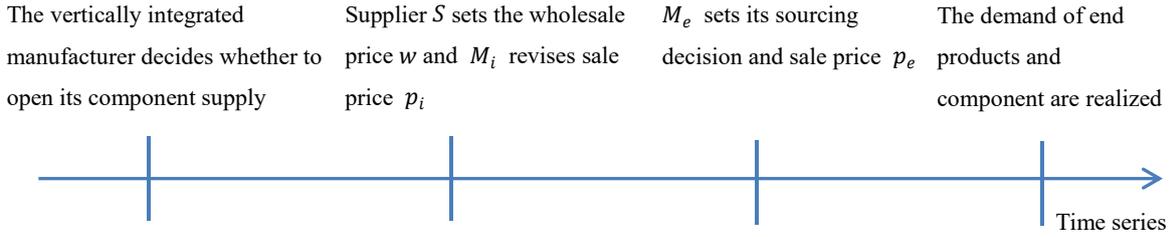

Fig. 2. Time series of events

### 3.2. Demand functions

Our demand model is inspired by a model originally introduced by Chiang, Chhajed, and Hess (2003). Initially, when the vertically integrated manufacturer is self-sufficiency, let $p_i^0$ denote its original product price. Let $p_i$ and $p_e$ denote the product price of $M_i$ and $M_e$ in the common market when the component supply is open. A customer would consider purchasing from $M_i$ if $v \geq p_i$, and would consider purchasing from $M_e$ if $\theta v \geq p_e$. When $v \geq p_i$ and $\theta v \geq p_e$, the customer chooses the product that remains the most surplus. That is, a customer buys from $M_i$ if $v \geq p_i$ and $v - p_i \geq \theta v - p_e$, and buys from $M_e$ if $\theta v \geq p_e$ and $\theta v - p_e \geq v - p_i$. We derive the product demand $Q_i^0 = A(1 - p_i^0)$ when the vertically integrated manufacturer is self-sufficiency. Following we derive the demand functions of $S$, $M_i$, and $M_e$ in scenarios $0 < \theta < 1$ and $\theta > 1$ when the vertically integrated manufacturer opens its component supply.

When $0 < \theta < 1$, the demand functions of $M_i$ and $M_e$ are given by

$$Q_i(p_i, p_e) = \begin{cases} 0 & p_i \geq 1 - \theta + p_e, \\ \hat{A}\left(1 - \frac{p_i - p_e}{1 - \theta}\right) & \frac{p_e}{\theta} \leq p_i < 1 - \theta + p_e, \\ \hat{A}(1 - p_i) & p_i < \frac{p_e}{\theta}. \end{cases} \quad (1)$$

and

$$Q_e(p_i, p_e) = \begin{cases} \hat{A}(1 - \frac{p_e}{\theta}) & p_i \geq 1 - \theta + p_e, \\ \hat{A}\frac{\theta p_i - p_e}{\theta(1-\theta)} & \frac{p_e}{\theta} \leq p_i < 1 - \theta + p_e, \\ 0 & p_i < \frac{p_e}{\theta}. \end{cases} \quad (2)$$

The component demand of $S$ is the sum of the product demand of $M_i$ and $M_e$, which can be formulated as follows:

$$Q_s(p_i, p_e) = \begin{cases} \hat{A}(1 - \frac{p_e}{\theta}) & p_i \geq \frac{p_e}{\theta}, \\ \hat{A}(1 - p_i) & p_i < \frac{p_e}{\theta}. \end{cases} \quad (3)$$

When $\theta > 1$, the demand functions of $M_i$ and $M_e$ are expressed by

$$Q_i(p_i, p_e) = \begin{cases} 0 & p_i > \frac{p_e}{\theta}, \\ \hat{A}\frac{p_e - \theta p_i}{\theta - 1} & p_e - \theta + 1 < p_i \leq \frac{p_e}{\theta}, \\ \hat{A}(1 - p_i) & p_i \leq p_e - \theta + 1. \end{cases} \quad (4)$$

and

$$Q_e(p_i, p_e) = \begin{cases} \hat{A}(1 - \frac{p_e}{\theta}) & p_i > \frac{p_e}{\theta}, \\ \hat{A}(1 - \frac{p_e - p_i}{\theta - 1}) & p_e - \theta + 1 < p_i \leq \frac{p_e}{\theta}, \\ 0 & p_i \leq p_e - \theta + 1. \end{cases}$$

(5)

Then, we derive the component demand function of $S$ in this scenario as follows:

$$Q_s(p_i, p_e) = \begin{cases} \hat{A}\left(1 - \frac{p_e}{\theta}\right) & p_i > p_e - \theta + 1, \\ \hat{A}(1 - p_i) & p_i \leq p_e - \theta + 1. \end{cases}$$

(6)

## 4. Model analysis

Assume that each player knows the perfect information before making decisions. To make sure that the game is subgame perfect, we first analyze the exterior manufacturer's decision in the third stage, then explore the pricing decisions of the vertically integrated manufacturer in the second stage, and at last examine the vertically integrated manufacturer's supply strategy choice in the first stage.

### *4.1. The exterior manufacturer's sourcing and pricing problem*

In this stage, $M_e$ needs to weigh the benefits of sourcing the newly opened component from $S$ against the original benefits of cooperating with its initial component supplier. Let $w_0$ and $p_e^0$ denote the purchase price from the initial component supplier and the product price of $M_e$. Let $\gamma_2(1-A)$ and $\pi_{M_e}^0$ denote the market share and profits of $M_e$ in open-supply context. Then we can derive the initial optimal sale price $p_e^{0*} = \frac{1+w_0+m_e}{2}$ and optimal profits $\pi_{M_e}^{0*} = \gamma_2(1-A)(\frac{1-w_0-m_e}{2})^2$ by solving the maximization problem

Max $(p_e^0 - w_0 - m_e)[\gamma_2(1-A)(1-p_e^0)]$.

$M_e$ does not make any decision if the vertically integrated manufacturer closes its component supply. Thus, this subsection only discusses the open-supply context. According to the demand models in SubSection 3.2, the sourcing and pricing decisions of $M_e$ is divided into the following two scenarios.

**Scenario I:** $0 < \theta < 1$. The profit for $M_e$ is given by

$$\pi_{M_e,3}(p_i,p_e,w) = \begin{cases} \hat{A}\left(1-\frac{p_e}{\theta}\right)(p_e-w-m_e) & p_e \leq p_i - 1 + \theta, \\ \hat{A}\frac{\theta p_i - p_e}{\theta(1-\theta)}(p_e-w-m_e) & p_i - 1 + \theta < p_e \leq \theta p_i, \\ 0 & p_e > \theta p_i. \end{cases} \quad (7)$$

**Proposition 1 (Best response of the exterior manufacturer in scenario $0 < \theta < 1$).** *Given the interior manufacturer's end product sale price $p_i$ and supplier's wholesale price w, the optimal sourcing decision and the equilibrium sale price of the exterior manufacturer are*:

i. *If $(w,p_i) \in R_1, R_2$ or $R_3$, the exterior manufacturer will choose to procure from supplier S and set its sale price as*

$$p_e^* = \begin{cases} \frac{\theta+w+m_e}{2} & (p_i,w) \in R_1, \\ \frac{\theta p_i+w+m_e}{2} & (p_i,w) \in R_2, \\ p_i - 1 + \theta & (p_i,w) \in R_3, \end{cases} \quad (8)$$

where $R_1 = \left\{(p_i,w) \,\middle|\, \begin{array}{l} w \leq 2p_i - 2 + \theta - m_e, \\ w \leq p_i - m_e, \\ w \leq \theta - m_e - \sqrt{\frac{4\theta\pi_{M_e}^{0*}}{\hat{A}}} \end{array}\right\}$,

$R_2 = \left\{(p_i,w) \,\middle|\, \begin{array}{l} w \geq (2-\theta)p_i - 2 + 2\theta - m_e, \\ w \leq p_i - m_e, \\ w \nleq p_i - m_e - \sqrt{\frac{4\theta(1-\theta)\pi_{M_e}^{0*}}{\hat{A}}} \end{array}\right\}$,

$R_3 = \left\{(p_i,w) \,\middle|\, \begin{array}{l} 2p_i - 2 + \theta - m_e \leq w \leq (2-\theta)p_i - 2 + 2\theta - m_e, \\ w \leq p_i - m_e, \\ \frac{\hat{A}(1-p_i)}{\theta}(p_i - 1 + \theta - w - m_e) \geq \pi_{M_e}^{0*} \end{array}\right\}$.

ii. *Otherwise, the exterior manufacturer would not be attracted to supplier S.*

The first inequality constraints in $R_1$, $R_2$ and $R_3$ are derived from the requirement for the piecewise-linear demand function $p_e^* \leq p_i - 1 + \theta$, $p_i - 1 + \theta \leq p_e^* \leq \theta p_i$, and $p_e^* = p_i - 1 + \theta$, respectively. The second and the third inequality constraints in the above three regions are derived from the prevention that $M_e$ buys the end product directly from $M_i$ (i.e., $w \leq p_i - m_e$) and the purpose that $M_e$ pursues higher profits than cooperating with the initial supplier (i.e., $\pi^*_{M_e,3} \geq \pi^{0*}_{M_e}$). In region $R_2$, the two manufacturers share the common market. In regions $R_1$ and $R_3$, the product price of $M_i$ is sufficiently high, which encourages $M_e$ to enter the common market and capture the whole market. The three regions are illustrated in Fig. 3a.

Proposition 1 provides the optimal sourcing decision and the equilibrium sale price of $M_e$ in scenario $0 < \theta < 1$. A high product sale price of $M_i$ (i.e., in region $R_1$) will not affect the price decision of $M_e$. When both manufacturers compete in the common market (i.e., in region $R_2$), the optimal price of $M_e$ will be influenced by the decision combination of $(p_i, w)$. Moreover, a moderate product sale price of $M_i$ (i.e., in region $R_3$) leads $M_e$ to ignore the impact of $w$.

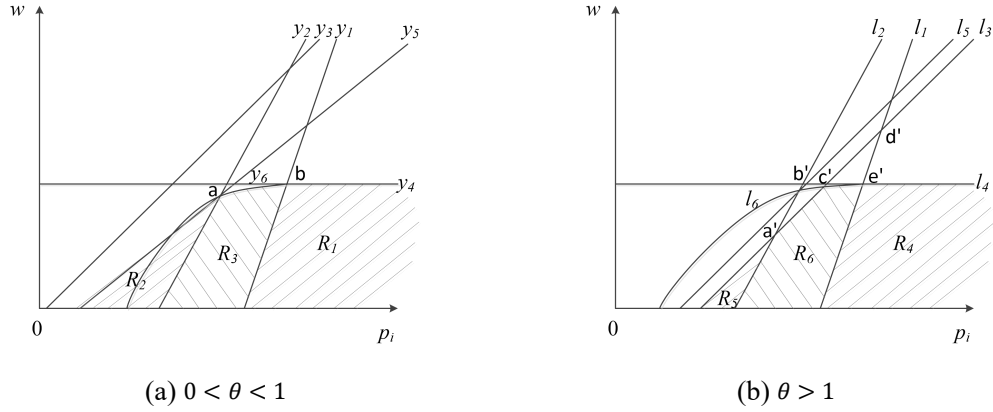

(a) $0 < \theta < 1$    (b) $\theta > 1$

Note 1: In Fig. 3a, line $y_1: w = 2p_i - 2 + \theta - m_e$; $y_2: w = (2-\theta)p_i - 2(1-\theta) - m_e$; $y_3: w = p_i - m_e$; $y_4: w = \theta - m_e - 2\sqrt{\frac{\theta \pi^{0*}_{M_e}}{\hat{A}}}$; $y_5: w = \theta p_i - m_e - 2\sqrt{\frac{\theta(1-\theta)\pi^{0*}_{M_e}}{\hat{A}}}$; $y_6: w = p_i - 1 + \theta - m_e - \frac{\theta \pi^{0*}_{M_e}}{\hat{A}(1-p_i)}$. The intersection of $y_2$, $y_5$ and $y_6$ is $a\left(1 - \sqrt{\frac{\theta \pi^{0*}_{M_e}}{\hat{A}(1-\theta)}}, \theta - m_e - (2-\theta)\sqrt{\frac{\theta \pi^{0*}_{M_e}}{\hat{A}(1-\theta)}}\right)$. The intersection of $y_1$, $y_4$ and $y_6$ is $b\left(1 - \sqrt{\frac{\theta \pi^{0*}_{M_e}}{\hat{A}}}, \theta - m_e - 2\sqrt{\frac{\theta \pi^{0*}_{M_e}}{\hat{A}}}\right)$.

Note 2: In Fig. 3b, line $l_1: w = 2\theta p_i - \theta - m_e$; $l_2: w = (2\theta - 1)p_i - \theta + 1 - m_e$; $l_3: w = p_i - m_e$; $l_4: w = \theta - m_e - 2\sqrt{\frac{\theta \pi^{0*}_{M_e}}{\hat{A}}}$; $l_5: w = p_i + \theta - 1 - m_e - 2\sqrt{\frac{(\theta-1)\pi^{0*}_{M_e}}{\hat{A}}}$; $l_6: w = \theta p_i - m_e - \frac{\pi^{0*}_{M_e}}{\hat{A}(1-p_i)}$. The intersection of $l_2$ and $l_3$, $l_2$ and $l_5$, $l_3$ and $l_6$, $l_1$ and $l_3$, $l_1$, $l_4$ and $l_6$ are $a'\left(\frac{1}{2}, \frac{1}{2} - m_e\right)$, $b'\left(1 - \sqrt{\frac{\pi^{0*}_{M_e}}{\hat{A}(\theta-1)}}, \theta - m_e - (2\theta - 1)\sqrt{\frac{\pi^{0*}_{M_e}}{\hat{A}(\theta-1)}}\right)$, $c'\left(\left[1 + \sqrt{1 - \frac{4\pi^{0*}_{M_e}}{\hat{A}(\theta-1)}}\right]/2, \left[1 + \sqrt{1 - \frac{4\pi^{0*}_{M_e}}{\hat{A}(\theta-1)}}\right]/2 - m_e\right)$, $d'\left(\frac{\theta}{2\theta-1}, \frac{\theta}{2\theta-1} - m_e\right)$ and $e'\left(1 - \sqrt{\frac{\pi^{0*}_{M_e}}{\theta \hat{A}}}, \theta - m_e - 2\sqrt{\frac{\theta \pi^{0*}_{M_e}}{\hat{A}}}\right)$.

Fig. 3. Feasible regions for $(p_i, w)$ when $0 < \theta < 1$ and $\theta > 1$

**Proposition 2:** ① If $\hat{A} \geq \frac{\theta(2-\theta)^2 \pi^{0*}_{M_e}}{(1-\theta)(\theta - m_e)^2}$, then $R_1, R_2, R_3 \neq \emptyset$; ② If $\frac{4\theta \pi^{0*}_{M_e}}{(\theta - m_e)^2} \leq \hat{A} < \frac{\theta(2-\theta)^2 \pi^{0*}_{M_e}}{(1-\theta)(\theta - m_e)^2}$, then $R_2 =$ $\emptyset, R_1, R_3 \neq \emptyset$; ③ If $\hat{A} < \frac{4\theta\pi_{M_e}^{0*}}{(\theta - m_e)^2}$, then $R_1, R_2, R_3 = \emptyset$.

Proposition 2 illustrates how $m_e$, $\pi_{M_e}^{0*}$, $\theta$, and $\hat{A}$ affect $M_e$'s decision when customers have a brand preference to $M_i$. If $\hat{A}$ is greater than the threshold $\frac{\theta(2-\theta)^2\pi_{M_e}^{0*}}{(1-\theta)(\theta-m_e)^2}$, $M_e$ will procure from $S$ when $(p_i, w)$ are set in the above three feasible regions. If $\hat{A}$ is moderate, setting a moderate or high $p_i$ in regions $R_1$ and $R_3$ is wiser for the vertically integrated manufacturer to capture component markets. However, a small $\hat{A}$ leads $M_e$ not to join the market even if the component is free.

**Scenario II:** $\theta > 1$. The profit for $M_e$ is:

$$\pi_{M_e,3}(p_i, p_e, w) = \begin{cases} \hat{A}\left(1 - \frac{p_e}{\theta}\right)(p_e - w - m_e) & p_e < \theta p_i, \\ \hat{A}\left(1 - \frac{p_e - p_i}{\theta - 1}\right)(p_e - w - m_e) & \theta p_i \leq p_e < p_i + \theta - 1, \\ 0 & p_e \geq p_i + \theta - 1. \end{cases} \quad (9)$$

**Proposition 3 (Best response of the exterior manufacturer in scenario $\theta > 1$).** *Given the interior manufacturer's end product sale price $p_i$ and supplier's wholesale price $w$, the optimal sourcing decision and the equilibrium sale price of the exterior manufacturer are:*

i. *If $(w, p_i) \in R_4, R_5$ or $R_6$, the exterior manufacturer will choose to procure from the supplier $S$ and set its sale price as*

$$p_e^* = \begin{cases} \frac{\theta + w + m_e}{2} & (p_i, w) \in R_4, \\ \frac{\theta - 1 + p_i + w + m_e}{2} & (p_i, w) \in R_5, \\ \theta p_i & (p_i, w) \in R_6, \end{cases} \quad (10)$$

where $R_4 = \left\{(p_i, w) \middle| \begin{array}{c} w < 2\theta p_i - \theta - m_e, \\ w \leq p_i - m_e, \\ w \leq \theta - m_e - \sqrt{\frac{4\theta\pi_{M_e}^{0*}}{\hat{A}}} \end{array} \right\}$,

$R_5 = \left\{(p_i, w) \middle| \begin{array}{c} w \geq (2\theta - 1)p_i - \theta + 1 - m_e, \\ w \leq p_i - m_e, \\ w \leq p_i - m_e + \theta - 1 - \sqrt{\frac{4(\theta-1)\pi_{M_e}^{0*}}{\hat{A}}} \end{array}\right\}$,

$R_6 = \left\{(p_i, w) \middle| \begin{array}{c} 2\theta p_i - \theta - m_e \leq w \leq (2\theta - 1)p_i - \theta + 1 - m_e, \\ w \leq p_i - m_e, \\ \hat{A}(1 - p_i)(\theta p_i - w - m_e) \geq \pi_{M_e}^{0*} \end{array}\right\}$.

ii. *Otherwise, the exterior manufacturer would not be attracted to supplier S.*

Proposition 3 presents that the optimal sale price of $M_e$ is also categorized into three regions in scenario $\theta > 1$. In region $R_5$, the two manufacturers share the common market. In regions $R_4$ and $R_6$, $M_e$ captures the whole common market. The three possible regions are illustrated in Fig. 3b. Both Fig. 3a and Fig. 3b demonstrate that the necessary conditions for $M_e$ to change a component supplier include an upper bound on $w$ and a lower bound on $p_i$.

**Proposition 4:** ① If $\hat{A} \geq \frac{(2\theta-1)^2 \pi_{M_e}^{0*}}{(\theta-1)(\theta-m_e)^2}$ and $m_e \leq \frac{1}{2}$, then $R_4, R_5, R_6 \neq \emptyset$; ② If $\hat{A} \geq \frac{4\theta\pi_{M_e}^{0*}}{(\theta-m_e)^2}$ and $\frac{1}{2} < m_e \leq \frac{\theta}{2\theta-1}$ or $\frac{4\theta\pi_{M_e}^{0*}}{(\theta-m_e)^2} \leq \hat{A} < \frac{(2\theta-1)^2\pi_{M_e}^{0*}}{(\theta-1)(\theta-m_e)^2}$ and $m_e \leq \frac{1}{2}$, then $R_5 = \emptyset, R_4, R_6 \neq \emptyset$; ③ If $\hat{A} \geq \frac{4\theta\pi_{M_e}^{0*}}{(\theta-m_e)^2}$ and $m_e > \frac{\theta}{2\theta-1}$, then $R_5, R_6 = \emptyset, R_4 \neq \emptyset$; ④ If $\hat{A} < \frac{4\theta\pi_{M_e}^{0*}}{(\theta-m_e)^2}$, then $R_4, R_5, R_6 = \emptyset$.

The first case in Proposition 4 illustrates that $M_e$ with low operation cost will benefit from entering a large common market. The second case shows that $M_e$ with moderate operation cost will monopoly the common market provided that $\hat{A}$ is not too small. If $m_e$ is relatively low, $M_e$ will occupy the whole market when $\hat{A}$ is at a moderate level. Furthermore, in the third case, when $m_e$ is relatively high and $\hat{A}$ is not too small, $M_e$ will capture the whole market. As to the fourth case, $M_e$ will not enter the market if $\hat{A}$ is too small.

### *4.2. The vertically integrated manufacturer's wholesale pricing and sale pricing problem*

After predicting the sourcing and pricing decision of $M_e$, the problem of the vertically integrated manufacturer is to maximize its total profits by rationally setting the component wholesale price and product sale price in open-supply context. Let $\Pi_{M_i}^0$ denote the profits of $M_i$ in the original oligopoly market, and $\Pi_{M_{i+s},i}$ denote the profit of the vertically integrated manufacturer in stage $i$ in open-supply context. When closing the component supply, $M_i$ monopolizes its original market share. We can derive the optimal sale price $p_i^{0*} = \frac{1+m_i}{2}$ and optimal profits $\Pi_{M_i}^{0*} = A\left(\frac{1-m_i}{2}\right)^2$ by solving the problem Max $A(p_i^0 - m_i)(1 - p_i^0)$. Now we consider the open-supply context. The component open-supply strategy is a long term strategy that the vertically integrated manufacturer adopts after careful consideration. The failure to receive component orders from the exterior manufacturer can lead to a negative impact on the reputation of the vertically integrated manufacturer, which further reduces the customer perceived value to its end products. Thus, we assume that the vertically integrated manufacturer aims to enter the component market without losing interest when making the open-supply strategy. We do not consider the choice of becoming only a vendor of end products in open-supply context.

**Scenario I:** $0 < \theta < 1$. $M_i$ has better customer perception. When $(p_i, w)$ is set in region $R_1$ or $R_3$, $\Pi_{M_{i+s},2}$ is only from selling components. Using Proposition 1 and Eq. (3), we derive the optimal $(p_i, w)$ in regions $R_1$ and $R_3$ by solving the following problem:

$$\text{Max}_{(p_i,w)} \quad \frac{w\hat{A}(\theta-w-m_e)}{2\theta}, \tag{11}$$

$$\text{Max}_{(p_i,w)} \quad \frac{w\hat{A}(1-p_i)}{\theta}. \tag{12}$$

**Lemma 1:** *In region $R_1$, the end product sale price $p_i^*$ of the two cases is arbitrarily large only if $(p_i^*, w^*) \in R_1$, and ① if $\hat{A} \geq \frac{16\theta \pi_{M_e}^{0*}}{(\theta-m_e)^2}$, then the supplier sets the optimal wholesale price at $w_1^{R_1*} = \frac{\theta-m_e}{2}$; ② If $\frac{4\theta \pi_{M_e}^{0*}}{(\theta-m_e)^2} \leq \hat{A} < \frac{16\theta \pi_{M_e}^{0*}}{(\theta-m_e)^2}$, then the supplier sets the optimal wholesale price at $w_2^{R_1*} = \theta - m_e - 2\sqrt{\frac{\theta \pi_{M_e}^{0*}}{\hat{A}}}$.*

In practice, the difference in customer perceived value between $M_i$ and $M_e$ is not particularly significant. Therefore, we assume that $\theta$ is large enough to ensure that the solutions are interior solutions.

From Proposition 2, we know that $R_1 \neq \emptyset$ if $\hat{A} \geq \frac{4\theta \pi_{M_e}^{0*}}{(\theta-m_e)^2}$ and is constrained by line $y_1$ and $y_4$ as Fig. 3a shows. Lemma 1 indicates that the variation of $p_i^*$ does not affect $\Pi_{M_{i+s},2}$ in region $R_1$. It implies that we can exclude region $R_1$ from the optimal pricing strategy, because the optimal prices in region $R_1$ are also in region $R_3$ and regions $R_1$ and $R_3$ are both nonempty when $\hat{A} \geq \frac{4\theta \pi_{M_e}^{0*}}{(\theta-m_e)^2}$ (see proof of Lemma 1 and Proposition 2).

**Lemma 2:** *In region $R_3$, ① if $\hat{A} \geq \frac{4\theta(2-\theta)^2 \pi_{M_e}^{0*}}{(1-\theta)(\theta-m_e)^2}$, then the optimal end product sale price and component wholesale price of the vertically integrated manufacturer are $p_{i1}^{R_3*} = \frac{4-3\theta+m_e}{2(2-\theta)}$, $w_1^{R_3*} = \frac{\theta-m_e}{2}$; ② If $\frac{4\theta \pi_{M_e}^{0*}}{(1-\theta)(\theta-m_e)^2} \leq \hat{A} < \frac{4\theta(2-\theta)^2 \pi_{M_e}^{0*}}{(1-\theta)(\theta-m_e)^2}$, then $p_{i2}^{R_3*} = 1 - \sqrt{\frac{\theta \pi_{M_e}^{0*}}{\hat{A}(1-\theta)}}$, $w_2^{R_3*} = \theta - m_e - (2-\theta)\sqrt{\frac{\theta \pi_{M_e}^{0*}}{\hat{A}(1-\theta)}}$; ③ If $\frac{4\theta \pi_{M_e}^{0*}}{(\theta-m_e)^2} \leq \hat{A} < \frac{4\theta \pi_{M_e}^{0*}}{(1-\theta)(\theta-m_e)^2}$, then $p_{i3}^{R_3*} = \frac{2-\theta+m_e}{2}$, $w_3^{R_3*} = \frac{\theta-m_e}{2} - \frac{2\theta \pi_{M_e}^{0*}}{\hat{A}(\theta-m_e)}$.*

Region $R_3$ is nonempty if $\hat{A} \geq \frac{4\theta \pi_{M_e}^{0*}}{(\theta-m_e)^2}$ as presented in Proposition 2 and is constrained by line $y_1$, $y_2$ and $y_6$ as illustrated in Fig. 3a. Lemma 2 shows that the optimal solutions in region $R_3$ are all boundary solutions (line $y_2$, point $a$, and curve $y_6$). Thus, the solutions on line $y_1$ cannot be optimal in region $R_3$, which implies that the optimal solutions in region $R_1$ are profit dominated by that in region $R_3$ when $R_1, R_3 \neq \emptyset$.

In region $R_2$, $\Pi_{M_{i+s},2}$ may come from two parts: selling components and end products. According to Proposition 1 and Eqs. (1) and (2), we can derive the optimal $(p_i, w)$ by solving the following problem:

$$\text{Max}_{(p_i,w)} \quad \hat{A}(p_i - m_i)\left[1 - \frac{(2-\theta)p_i - w - m_e}{2(1-\theta)}\right] + \frac{w\hat{A}(\theta p_i - w - m_e)}{2\theta(1-\theta)}. \tag{13}$$

**Lemma 3:** *In region $R_2$, ①if $\hat{A} \geq \frac{16\theta(1-\theta)\pi_{M_e}^{0*}}{(\theta m_i - m_e)^2}$ and $1 - m_i \geq \frac{\theta - m_e}{2-\theta}$, then the optimal end product sale price and component wholesale price of the vertically integrated manufacturer are $p_{i1}^{R_2*} = \frac{1+m_i}{2}$, $w_1^{R_2*} = \frac{\theta-m_e}{2}$; ②If $\hat{A} \geq \frac{4\theta(2-\theta)^2\pi_{M_e}^{0*}}{(1-\theta)(\theta-m_e)^2}$ and $1 - m_i \leq \frac{\theta-m_e}{2-\theta}$, then $(p_{i2}^{R_2*}, w_2^{R_2*}) = (p_{i1}^{R_3*}, w_1^{R_3*})$; ③If $\frac{4\theta\pi_{M_e}^{0*}}{(1-m_i)^2(1-\theta)} \leq \hat{A} < \frac{16\theta(1-\theta)\pi_{M_e}^{0*}}{(\theta m_i - m_e)^2}$, then $p_{i3}^{R_2*} = \frac{1+m_i}{2}$, $w_3^{R_2*} = \frac{\theta(1+m_i)}{2} - m_e - 2\sqrt{\frac{\theta(1-\theta)\pi_{M_e}^{0*}}{\hat{A}}}$; ④ If $\hat{A} < \min\left\{\frac{4\theta\pi_{M_e}^{0*}}{(1-m_i)^2(1-\theta)}, \frac{4\theta(2-\theta)^2\pi_{M_e}^{0*}}{(1-\theta)(\theta-m_e)^2}\right\}$, then $(p_{i4}^{R_2*}, w_4^{R_2*}) = (p_{i2}^{R_3*}, w_2^{R_3*})$.*

From Proposition 2, region $R_2$ is nonempty if $\hat{A} \geq \frac{\theta(2-\theta)^2\pi_{M_e}^{0*}}{(1-\theta)(\theta-m_e)^2}$ and is subject to line $y_2$ and $y_5$ as illustrated in Fig. 3a. Lemma 3 shows that the optimal prices in the second and the fourth case lie on the common boundary of region $R_2$ and $R_3$ (i.e., line $y_2$), which are the same as that of the first and the second case in Lemma 2. As for the first case and the third case, the optimal prices are obtained in the interior of region $R_2$ and on line $y_5$ respectively, which implies that the two manufacturers share the common market. By comparing the optimal solutions among the above three subproblems, we derive the equilibrium outcomes in stage 2 when the vertically integrated manufacturer opens component supply.

**Proposition 5 (Equilibrium pricing decisions for the vertically integrated manufacturer).** *In scenario $0 < \theta < 1$, if the vertically integrated manufacturer opens the component supply, the optimal sale price $p_i^*$ and the optimal wholesale price $w^*$ are set in one of the following two mutually exclusive and collective exhaustive cases as shown in Table 2.*

Table 2. The Equilibrium Outcomes in Stage 2 in scenario $0 < \theta < 1$

| Cases | $p_i^*$ | $w^*$ | $\Pi_{M_{i+s,2}}$ |
|---|---|---|---|
| i $\hat{A} \geq \frac{\theta(2-\theta)^2\pi_{M_e}^{0*}}{(1-\theta)(\theta-m_e)^2}$ | | | |
| i.1 $1 - m_i \geq \frac{\theta - m_e}{2-\theta}$ | | | |
| i.1.1 $\hat{A} \geq \frac{16\theta(1-\theta)\pi_{M_e}^{0*}}{(\theta m_i - m_e)^2}$ | $p_{i1}^{R_2*}$ | $w_1^{R_2*}$ | $\Pi_{M_{i+s,2}}(p_{i1}^{R_2*}, w_1^{R_2*})$ |
| i.1.2 $\frac{4\theta\pi_{M_e}^{0*}}{(1-m_i)^2(1-\theta)} \leq \hat{A} < \frac{16\theta(1-\theta)\pi_{M_e}^{0*}}{(\theta m_i - m_e)^2}$ | | | |
| i.1.2.1 $\hat{A} \geq \hat{A}^0(m_i, m_e, \theta)$ (if applicable) | $p_{i3}^{R_2*}$ | $w_3^{R_2*}$ | $\Pi_{M_{i+s,2}}(p_{i3}^{R_2*}, w_3^{R_2*})$ |
| i.1.2.2 $\hat{A} < \hat{A}^0(m_i, m_e, \theta)$ (if applicable) | $p_{i3}^{R_3*}$ | $w_3^{R_3*}$ | $\Pi_{M_{i+s,2}}(p_{i3}^{R_3*}, w_3^{R_3*})$ |
| i.1.3 $\frac{4\theta\pi_{M_e}^{0*}}{(1-\theta)(\theta-m_e)^2} < \hat{A} < \frac{4\theta\pi_{M_e}^{0*}}{(1-m_i)^2(1-\theta)}$ | $p_{i2}^{R_3*}$ | $w_2^{R_3*}$ | $\Pi_{M_{i+s,2}}(p_{i2}^{R_3*}, w_2^{R_3*})$ |
| i.1.4 $\frac{4\theta\pi_{M_e}^{0*}}{(\theta-m_e)^2} \leq \hat{A} \leq \frac{4\theta\pi_{M_e}^{0*}}{(1-\theta)(\theta-m_e)^2}$ | $p_{i3}^{R_3*}$ | $w_3^{R_3*}$ | $\Pi_{M_{i+s,2}}(p_{i3}^{R_3*}, w_3^{R_3*})$ |

| | | | | |
|---|---|---|---|---|
| i.2 $1-m_i < \frac{\theta-m_e}{2-\theta}$ | | | | |
| i.2.1 $\hat{A} \geq \frac{4\theta(2-\theta)^2 \pi_{M_e}^{0*}}{(1-\theta)(\theta-m_e)^2}$ | $p_{i1}^{R_3 *}$ | $w_1^{R_3 *}$ | $\Pi_{M_{i+s},2}(p_{i1}^{R_3 *}, w_1^{R_3 *})$ | |
| i.2.2 $\frac{4\theta \pi_{M_e}^{0*}}{(1-\theta)(\theta-m_e)^2} < \hat{A} < \frac{4\theta(2-\theta)^2 \pi_{M_e}^{0*}}{(1-\theta)(\theta-m_e)^2}$ | $p_{i2}^{R_3 *}$ | $w_2^{R_3 *}$ | $\Pi_{M_{i+s},2}(p_{i2}^{R_3 *}, w_2^{R_3 *})$ | |
| i.2.3 $\frac{4\theta \pi_{M_e}^{0*}}{(\theta-m_e)^2} \leq \hat{A} \leq \frac{4\theta \pi_{M_e}^{0*}}{(1-\theta)(\theta-m_e)^2}$ | $p_{i3}^{R_3 *}$ | $w_3^{R_3 *}$ | $\Pi_{M_{i+s},2}(p_{i3}^{R_3 *}, w_3^{R_3 *})$ | |
| ii $\frac{4\theta \pi_{M_e}^{0*}}{(\theta-m_e)^2} \leq \hat{A} < \frac{\theta(2-\theta)^2 \pi_{M_e}^{0*}}{(1-\theta)(\theta-m_e)^2}$ | $p_{i3}^{R_3 *}$ | $w_3^{R_3 *}$ | $\Pi_{M_{i+s},2}(p_{i3}^{R_3 *}, w_3^{R_3 *})$ | |

Note: we define that $\hat{A}^0(m_i, m_e, \theta) \stackrel{\text{def}}{=} arg\{\Pi_{M_{i+s},2}(p_{i3}^{R_3 *}, w_3^{R_3 *}) - \Pi_{M_{i+s},2}(p_{i3}^{R_2 *}, w_3^{R_2 *}) = 0\}$.

Proposition 5 provides the equilibrium prices and profits of the vertically integrated manufacturer when $0 < \theta < 1$. The equilibrium decisions are divided into several subcases by $\hat{A}$, $m_i$, and $m_e$. In case i, the vertically integrated manufacturer may choose to be a "dual manufacturer" or a component manufacturer. There are two subcases i.1.1 and i.1.2.1 wherein the manufacturer chooses to be a "dual manufacturer" and sets the sale price equal to the initial closed-supply context. The two subcases happen when $m_i$ is not significantly larger than $m_e$ (subcase i.1 requires $m_i - m_e \leq \frac{(1-\theta)(2-m_e)}{2-\theta}$) and $\hat{A}$ is relatively large (subcase i.1.1 requires $\hat{A} \geq \frac{16\theta(1-\theta)\pi_{M_e}^{0*}}{(\theta m_i - m_e)^2}$ and subcase i.1.2.1 requires $\hat{A} \geq \hat{A}^0(m_i, m_e, \theta)$). Therefore, if the manufacturer wants to maintain the product manufacturing business and expand the component supply business, the decision-maker must consider whether the product market potential is large enough if opens the component supply, and whether its unit operation cost is not much larger than that of the exterior manufacturer even if its product can win more favor from customers. In case ii, i.e., $\hat{A}$ is at a medium level, the vertically integrated manufacturer becomes a component manufacturer in open-supply context.

**Scenario II:** $\theta > 1$. In this scenario, $M_i$ has weaker customer perception. When $(p_i, w)$ is set in region $R_4$ or $R_6$, $\Pi_{M_{i+s},2}$ is only from selling components. According to Proposition 2 and Eq. (6), we can derive the optimal $(p_i, w)$ in region $R_4$ and $R_6$ by solving the following problem respectively:

$$\text{Max}_{(p_i,w)} \quad \frac{w\hat{A}(\theta-w-m_e)}{2\theta}, \tag{14}$$

$$\text{Max}_{(p_i,w)} \quad w\hat{A}(1-p_i). \tag{15}$$

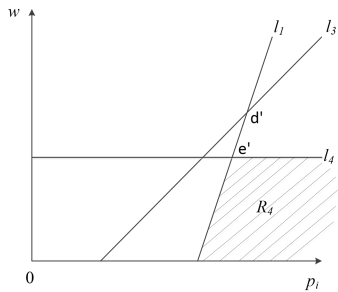 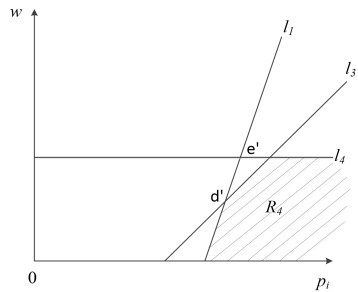

Fig. 4. Region $R_4$ for $(p_i, w)$ when $\theta > 1$

**Lemma 4:** *In region $R_4$, the end product sale price $p_i^*$ of the two cases is arbitrarily large only if $(p_i^*, w^*) \in R_4$, and ①if $\hat{A} \geq \frac{16\theta \pi_{M_e}^{0*}}{(\theta - m_e)^2}$, then the supplier sets the optimal wholesale price at $w_1^{R_4 *} = \frac{\theta - m_e}{2}$; ②If $\frac{4\theta \pi_{M_e}^{0*}}{(\theta - m_e)^2} \leq \hat{A} < \frac{16\theta \pi_{M_e}^{0*}}{(\theta - m_e)^2}$, then the supplier sets the optimal wholesale price at $w_2^{R_4 *} = \theta - m_e - 2\sqrt{\frac{\theta \pi_{M_e}^{0*}}{\hat{A}}}$.*

As Proposition 4 indicates, region $R_4$ is nonempty if $\hat{A} \geq \frac{4\theta \pi_{M_e}^{0*}}{(\theta - m_e)^2}$ and the possible cases of region $R_4$ are shown in Fig. 4. The optimal solutions in region $R_4$ are similar to that of region $R_1$. But the difference in $\theta$ and region scope makes the solution in regions $R_1$ and $R_4$ different. The optimal solution in region $R_4$ can be obtained on line $l_1$ when $\frac{4\theta \pi_{M_e}^{0*}}{(\theta - m_e)^2} \leq \hat{A} \leq \frac{(2\theta - 1)^2 \pi_{M_e}^{0*}}{\theta(\theta - 1)^2}$ as shown in Fig. 4a and be obtained on line $l_1$ or $l_3$ when $\hat{A} > \frac{(2\theta - 1)^2 \pi_{M_e}^{0*}}{\theta(\theta - 1)^2}$ as shown in Fig. 4b.

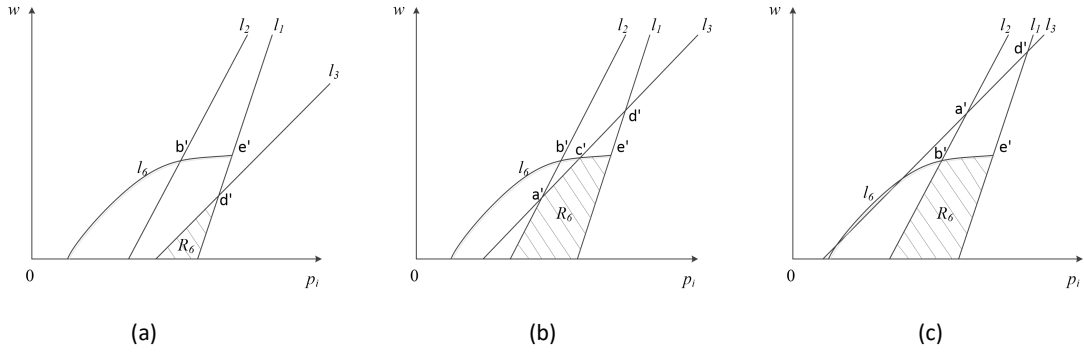

(a) (b) (c)

Fig. 5. Region $R_6$ for $(p_i, w)$ when $\theta > 1$

**Lemma 5:** *In region $R_6$, ①if $\hat{A} > \frac{(2\theta - 1)^2 \pi_{M_e}^{0*}}{\theta(\theta - 1)^2}$ and $\frac{1}{2\theta - 1} < m_e \leq \frac{\theta}{2\theta - 1}$, then the optimal end product sale price and component wholesale price of the vertically integrated manufacturer are $p_{i1}^{R_6 *} = \frac{\theta}{2\theta - 1}$, $w_1^{R_6 *} = \frac{\theta}{2\theta - 1} - m_e$; ②If $\hat{A} > \frac{4\pi_{M_e}^{0*}}{(\theta - 1)(1 - m_e^2)}$ and $m_e \leq \frac{1}{2\theta - 1}$, then $p_{i2}^{R_6 *} = \frac{1 + m_e}{2}$, $w_2^{R_6 *} = \frac{1 - m_e}{2}$; ③If $\frac{4\theta^2 \pi_{M_e}^{0*}}{(\theta - 1)(\theta^2 - m_e^2)} < \hat{A} \leq \frac{4\pi_{M_e}^{0*}}{(\theta - 1)(1 - m_e^2)}$, then the optimal prices are $p_{i3}^{R_6 *} = \frac{1}{2}\left[1 + \sqrt{1 - \frac{4\pi_{M_e}^{0*}}{\hat{A}(\theta - 1)}}\right]$, $w_3^{R_6 *} = \frac{1}{2}\left[1 + \sqrt{1 - \frac{4\pi_{M_e}^{0*}}{\hat{A}(\theta - 1)}}\right] - m_e$; ④If $\frac{4\theta \pi_{M_e}^{0*}}{(\theta - m_e)^2} \leq \hat{A} \leq \frac{4\theta^2 \pi_{M_e}^{0*}}{(\theta - 1)(\theta^2 - m_e^2)}$, then $p_{i4}^{R_6 *} = \frac{\theta + m_e}{2\theta}$, $w_4^{R_6 *} = \frac{\theta - m_e}{2} - \frac{2\theta \pi_{M_e}^{0*}}{\hat{A}(\theta - m_e)}$.*

Proposition 4 shows that if $\hat{A} \geq \frac{4\theta \pi_{M_e}^{0*}}{(\theta - m_e)^2}$ and $m_e \leq \frac{\theta}{2\theta - 1}$, region $R_6$ is nonempty. Moreover, region $R_6$ is constrained to line $l_1$, $l_2$ and $l_3$ when $\hat{A} > \frac{(2\theta - 1)^2 \pi_{M_e}^{0*}}{\theta(\theta - 1)^2}$ and is restricted to line $l_1$, $l_2$, $l_3$ and $l_6$ when

$\frac{4\pi_{M_e}^{0*}}{\theta-1} < \hat{A} \le \frac{(2\theta-1)^2 \pi_{M_e}^{0*}}{\theta(\theta-1)^2}$ as illustrated in Fig. 5a and Fig. 5b. Otherwise, when $\frac{4\theta\pi_{M_e}^{0*}}{(\theta-m_e)^2} \le \hat{A} \le \frac{4\pi_{M_e}^{0*}}{\theta-1}$, region $R_6$ is limited to line $l_1$, $l_2$ and $l_6$ as shown in Fig. 5c. In Lemma 5, the optimal solutions are obtained at point "d'" in Fig. 5a, on line $l_3$ in Fig. 5a or Fig. 5b, at point "c'" in Fig. 5b, and on line $l_6$ in Fig. 5b or Fig. 5c, respectively. We obtain that the first solution $(p_{i1}^{R_6*}, w_1^{R_6*})$ is also in region $R_4$ and is profit dominated by the optimal solutions in region $R_4$ (see the proof of Proposition 6).

When $(w, p_i) \in R_5$, $\Pi_{M_{i+s},2}$ may come from selling components and end products. Then we derive the optimal $(p_i, w)$ by solving the following problem:

$$\text{Max}_{(w,p_i)} \hat{A}(p_i - m_i)\left[\frac{1}{2} - \frac{(2\theta-1)p_i - w - m_e}{2(\theta-1)}\right] + w\hat{A}\left[\frac{1}{2} - \frac{w + m_e - p_i}{2(\theta-1)}\right]. \tag{16}$$

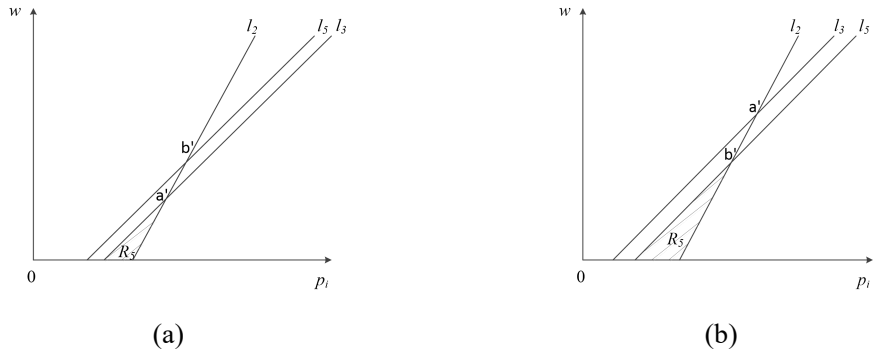

Fig. 6. Region $R_5$ for $(p_i, w)$ when $\theta > 1$

**Lemma 6:** *In region $R_5$, ① if $\hat{A} > \frac{4\pi_{M_e}^{0*}}{\theta-1}$ and $m_e \le \frac{1}{2}$, then the optimal end product sale price and component wholesale price of the vertically integrated manufacturer are $p_{i1}^{R_5*} = \frac{1}{2}$, $w_1^{R_5*} = \frac{1}{2} - m_e$; ② If $\frac{(2\theta-1)^2 \pi_{M_e}^{0*}}{(\theta-1)(\theta-m_e)^2} \le \hat{A} \le \frac{4\pi_{M_e}^{0*}}{\theta-1}$, then the optimal prices are $p_{i2}^{R_5*} = 1 - \sqrt{\frac{\pi_{M_e}^{0*}}{\hat{A}(1-\theta)}}$, $w_2^{R_5*} = \theta - m_e - (2\theta - 1)\sqrt{\frac{\pi_{M_e}^{0*}}{\hat{A}(1-\theta)}}$.*

From the proof of Proposition 4, region $R_5$ is constrained to line $l_2$ and $l_3$ when $\hat{A} > \frac{4\pi_{M_e}^{0*}}{\theta-1}$ and $m_e \le \frac{1}{2}$ and is constrained to line $l_2$ and $l_5$ when $\frac{(2\theta-1)^2 \pi_{M_e}^{0*}}{(\theta-1)(\theta-m_e)^2} \le \hat{A} \le \frac{4\pi_{M_e}^{0*}}{\theta-1}$ as shown in Fig. 6a and Fig. 6b. Lemma 6 shows that the optimal solution is always at their highest levels in region $R_5$, i.e., at point "$a'$" or point "$b'$". We can also derive that the prices at point "$a'$" and point "$b'$" are profit dominated by other prices in region $R_6$ (see proof of Lemma 5). How to price globally in scenario $\theta > 1$ is presented in Proposition 6.

**Proposition 6 (Equilibrium pricing decisions for the vertically integrated manufacturer).** *In scenario*

$\theta > 1$, *if the vertically integrated manufacturer opens the component supply, the optimal sale price* $p_i^*$ *and the optimal wholesale price* $w^*$ *are set in one of the following two mutually exclusive cases as shown in Table 3.*

Table 3. The Equilibrium Outcomes in Stage 2 in scenario $\theta > 1$

| Cases | $p_i^*$ | $w^*$ | $\Pi_{M_{i+s},2}$ |
|---|---|---|---|
| i $\hat{A} \geq \frac{4\theta\pi_{M_e}^{0*}}{(\theta-m_e)^2}$ and $m_e \leq \frac{\theta}{2\theta-1}$ | | | |
| i.1 $\hat{A} > \frac{(2\theta-1)^2\pi_{M_e}^{0*}}{\theta(\theta-1)^2}$ and $m_e > \frac{1}{2\theta-1}$ | | | |
| i.1.1 $\hat{A} \geq \frac{16\theta\pi_{M_e}^{0*}}{(\theta-m_e)^2}$ | Arbitrary in $R_4$ | $w_1^{R_4*}$ | $\Pi_{M_{i+s},2}(p_{i1}^{R_4*},w_1^{R_4*})$ |
| i.1.2 $\hat{A} < \frac{16\theta\pi_{M_e}^{0*}}{(\theta-m_e)^2}$ | Arbitrary in $R_4$ | $w_2^{R_4*}$ | $\Pi_{M_{i+s},2}(p_{i2}^{R_4*},w_2^{R_4*})$ |
| i.2 $\hat{A} > \frac{4\pi_{M_e}^{0*}}{(\theta-1)(1-m_e^2)}$ and $m_e \leq \frac{1}{2\theta-1}$ | | | |
| i.2.1 $\hat{A} \geq \frac{16\theta\pi_{M_e}^{0*}}{(\theta-m_e)^2}$ | Arbitrary in $R_4$ | $w_1^{R_4*}$ | $\Pi_{M_{i+s},2}(p_{i1}^{R_4*},w_1^{R_4*})$ |
| i.2.2 $\hat{A} < \frac{16\theta\pi_{M_e}^{0*}}{(\theta-m_e)^2}$ | | | |
| i.2.2.1 $\hat{A} \geq \hat{A}^1(m_i,m_e,\theta)$ (if applicable) | Arbitrary in $R_4$ | $w_2^{R_4*}$ | $\Pi_{M_{i+s},2}(p_{i2}^{R_4*},w_2^{R_4*})$ |
| i.2.2.2 $\hat{A} < \hat{A}^1(m_i,m_e,\theta)$ (if applicable) | $p_{i2}^{R_6*}$ | $w_2^{R_6*}$ | $\Pi_{M_{i+s},2}(p_{i2}^{R_6*},w_2^{R_6*})$ |
| i.3 $\frac{4\theta^2\pi_{M_e}^{0*}}{(\theta-1)(\theta^2-m_e^2)} < \hat{A} \leq \frac{4\pi_{M_e}^{0*}}{(\theta-1)(1-m_e^2)}$ | $p_{i3}^{R_6*}$ | $w_3^{R_6*}$ | $\Pi_{M_{i+s},2}(p_{i3}^{R_6*},w_3^{R_6*})$ |
| i.4 $\hat{A} \leq \frac{4\theta^2\pi_{M_e}^{0*}}{(\theta-1)(\theta^2-m_e^2)}$ | $p_{i4}^{R_6*}$ | $w_4^{R_6*}$ | $\Pi_{M_{i+s},2}(p_{i4}^{R_6*},w_4^{R_6*})$ |
| ii $\hat{A} \geq \frac{4\theta\pi_{M_e}^{0*}}{(\theta-m_e)^2}$ and $m_e > \frac{\theta}{2\theta-1}$ | | | |
| ii.1 $\hat{A} \geq \frac{16\theta\pi_{M_e}^{0*}}{(\theta-m_e)^2}$ | Arbitrary in $R_4$ | $w_1^{R_4*}$ | $\Pi_{M_{i+s},2}(p_{i1}^{R_4*},w_1^{R_4*})$ |
| ii.2 $\hat{A} < \frac{16\theta\pi_{M_e}^{0*}}{(\theta-m_e)^2}$ | Arbitrary in $R_4$ | $w_2^{R_4*}$ | $\Pi_{M_{i+s},2}(p_{i2}^{R_4*},w_2^{R_4*})$ |

Note: we define that $\hat{A}^1(m_i,m_e,\theta) \stackrel{\text{def}}{=} arg\{\Pi_{M_{i+s},2}(p_{i2}^{R_6*},w_2^{R_6*}) - \Pi_{M_{i+s},2}(p_{i2}^{R_4*},w_2^{R_4*}) = 0\}$.

Proposition 6 provides the equilibrium prices and profits of the vertically integrated manufacturer when $\theta > 1$. The equilibrium decisions are divided into several subcases by $\hat{A}$ and $m_e$. It indicates that the vertically integrated manufacturer with a weaker customer perceived value always chooses to be a component manufacturer in open-supply context. In case i, i.e., $\hat{A}$ is relatively large and $m_e$ is relatively low, the equilibrium solutions change from region $R_4$ to region $R_6$ with the contraction of $\hat{A}$. The optimal sale price $p_i^*$ is from an arbitrary value in a certain region to a fixed value when $\hat{A}$ shrinks. It implies that when $\hat{A}$ is large enough, setting an arbitrary price has no impact on the profit of the vertically integrated manufacturer provided the decision $(p_i^*, w^*)$ is within region $R_4$. Otherwise, as $\hat{A}$ narrows, the vertically integrated manufacturer must set a rational product sale price prevent $M_e$ from overpricing. In case ii, i.e., $\hat{A}$ is relatively large and $m_e$ is high, the equilibrium solutions are always in region $R_4$. It shows that if $m_e$ is high enough, the vertically integrated manufacturer can set an arbitrary sale price that guarantees $(p_i^*, w^*) \in R_4$ without affecting the profit of both manufacturers.

### 4.3. The vertically integrated manufacturer's supply strategy decision problem

The supply strategy decision aims to capture the component market and make more profits. Based on the equilibrium outcomes in SubSection 4.2, we subtract the investment amount and get the total profits of the vertically integrated manufacturer in this stage, i.e., $\Pi_{M_{i+s,1}}(p_i^*, w^*) = \Pi_{M_{i+s,2}}(p_i^*, w^*) - K$. If the profit $\Pi_{M_{i+s,1}}$ is surplus after subtracting the profits in closed-supply context, i.e., $\Pi_{M_{i+s,1}}(p_i^*, w^*) - \Pi_{M_i}^{0*} \geq 0$, the vertically integrated manufacturer will open component supply. For brevity, the component brand-building investment is assumed to be zero, i.e., $K = 0$. Due to the complicated features of the supply strategy's equilibrium outcome, we first present the comparison result when $\hat{A} < \frac{4\theta \pi_{M_e}^{0*}}{(\theta - m_e)^2}$, and the observations when $\hat{A} \geq \frac{4\theta \pi_{M_e}^{0*}}{(\theta - m_e)^2}$ are summarized through numerical examples.

**Proposition 7 (Generalized closed-supply Pareto zone).** *In scenarios $0 < \theta < 1$ and $\theta > 1$, the vertically integrated manufacturer closes its component supply with a small common market share where $\hat{A} < \frac{4\theta \pi_{M_e}^{0*}}{(\theta - m_e)^2}$.*

From Proposition 2 and Proposition 4, we know that when $\hat{A} < \frac{4\theta \pi_{M_e}^{0*}}{(\theta - m_e)^2}$, the vertically integrated manufacturer fails to attract any exterior manufacturer in scenarios $0 < \theta < 1$ and $\theta > 1$, thus the vertically integrated manufacturer closes the component supply. Moreover, we can derive $\partial \frac{4\theta \pi_{M_e}^{0*}}{(\theta - m_e)^2} / \partial \theta < 0$, which implies that the weaker the customer perceived value of the vertically integrated manufacturer, the narrower the generalized closed-supply Pareto zone.

Similar to Chen et al. (2017), to establish the supply strategy Pareto zone when $\hat{A} \geq \frac{4\theta \pi_{M_e}^{0*}}{(\theta - m_e)^2}$, we turn to a numerical example to verify the comparison results of the vertically integrated manufacturer's profits before and after the component open-supply strategy (see the last two columns of Table A1 and Table A2 in Appendix A). The default values of parameters in this subsection are $\gamma_2 = 0.5$ and $w_0 = 0.05$. We set $A$ and $m_e$ as the independent variable while fixing the value of other parameters. As $A$ or $m_e$ increases, $\pi_{M_e}^{0*}$ decreases. Based on the setting, we employ the inequality $\hat{A} \geq \frac{4\theta \pi_{M_e}^{0*}}{(\theta - m_e)^2}$ and $\gamma_1 + \gamma_2 \leq 1$ to constrain the parameter $\gamma_1$.

As shown in Table A1, the open-supply strategy provides the vertically integrated manufacturer and the whole system higher profits in addition to the setting of low original market share and weak market spillover in scenario $0 < \theta < 1$. Moreover, the vertically integrated manufacturer prefers to open component supply as a component manufacturer when $m_e$ is small. As shown in Table A2, a large original market hinders the vertically integrated manufacturer from opening the component supply. The open-supply strategy seems to be dominant under the setting of low original market share and strong market spillover in scenario $\theta > 1$. As $A$ becomes larger, the vertically integrated manufacturer prefers to close the component supply. We further identify the supply strategy Pareto zone under several configurations in Fig. 7.

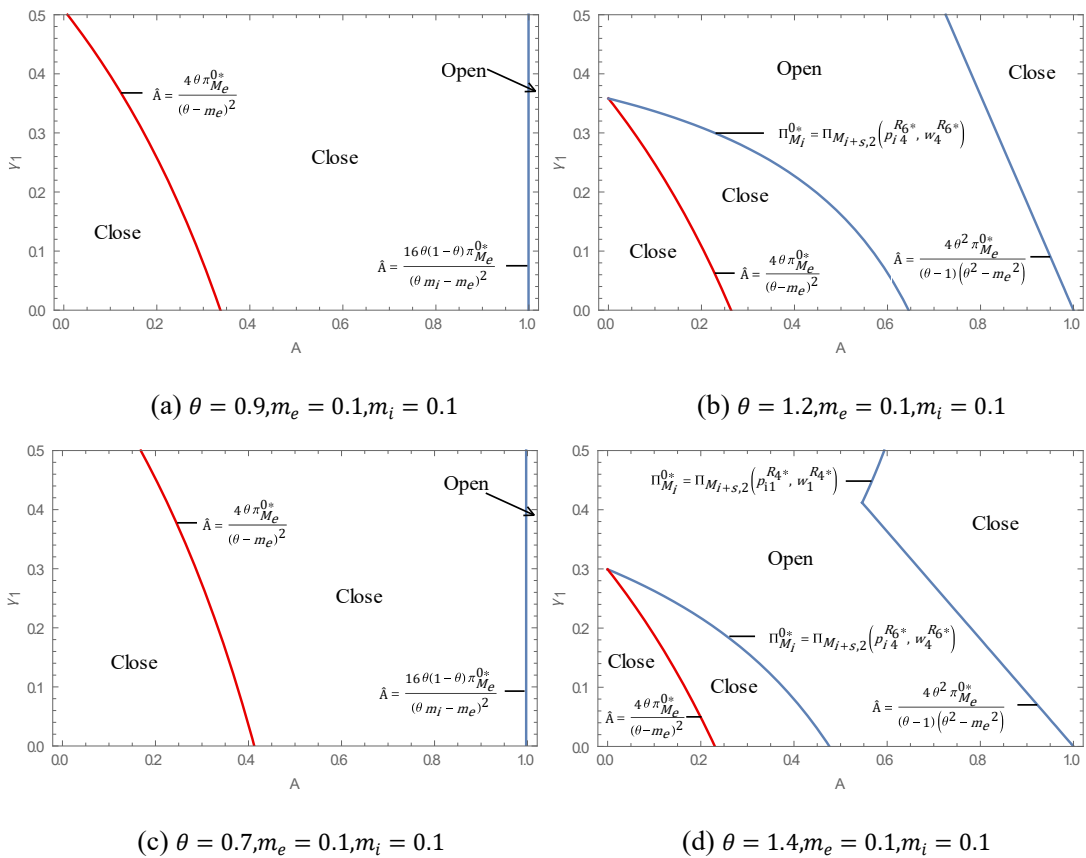

(a) $\theta = 0.9, m_e = 0.1, m_i = 0.1$

(b) $\theta = 1.2, m_e = 0.1, m_i = 0.1$

(c) $\theta = 0.7, m_e = 0.1, m_i = 0.1$

(d) $\theta = 1.4, m_e = 0.1, m_i = 0.1$

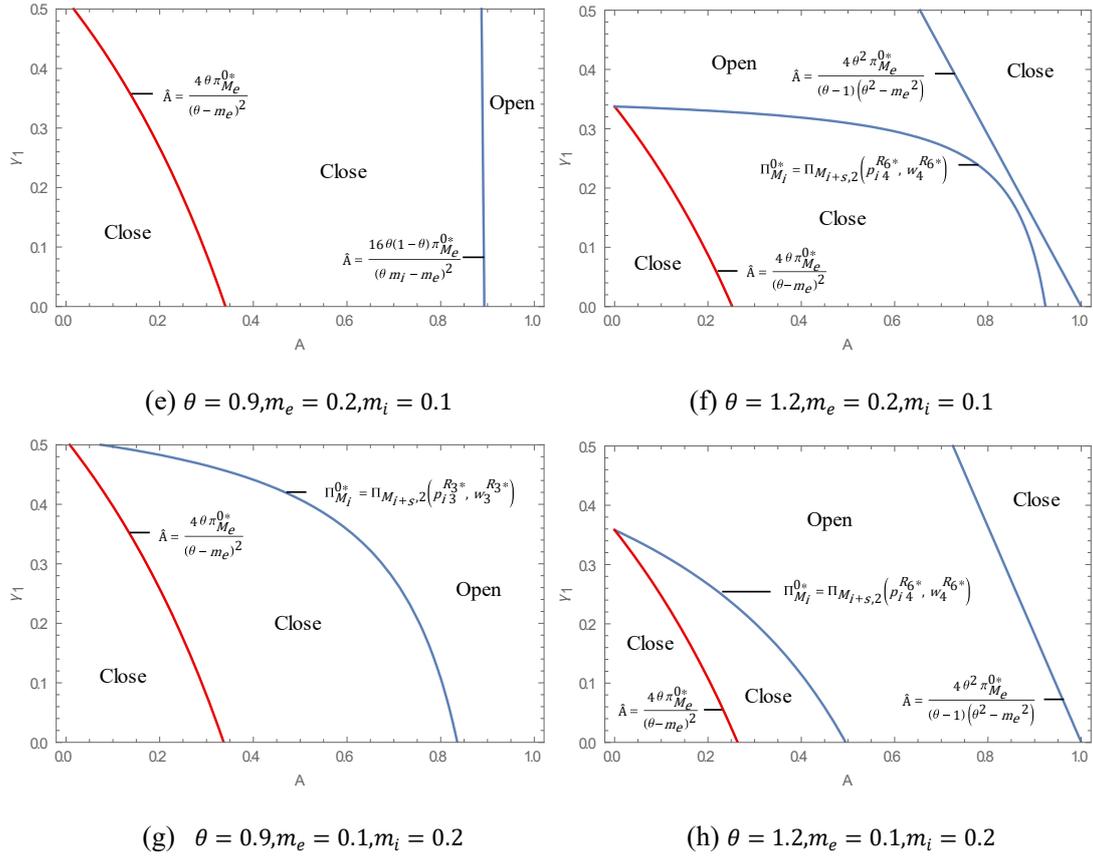

(e) $\theta = 0.9, m_e = 0.2, m_i = 0.1$

(f) $\theta = 1.2, m_e = 0.2, m_i = 0.1$

(g) $\theta = 0.9, m_e = 0.1, m_i = 0.2$

(h) $\theta = 1.2, m_e = 0.1, m_i = 0.2$

Fig. 7. The supply strategy Pareto zone

Realizing that market changes may influence the component supply strategy decision, we here examine how the original market share A and the market spillover $\gamma_1$ affect the equilibrium supply strategy. Fig. 7 confirms the conclusion in Proposition 7 that the generalized closed-supply Pareto zone is $\hat{A} < \frac{4\theta \pi_{M_e}^{0*}}{(\theta - m_e)^2}$ (outlined by the red curves). As the common market swells to $\hat{A} \geq \frac{4\theta \pi_{M_e}^{0*}}{(\theta - m_e)^2}$, we derive the following insights from the equilibrium supply strategy in Fig. 7 (divided by the blue curves):

(1) The vertically integrated manufacturer with better customer perception will close component supply unless it nearly monopolizes the product market initially or conducts a relatively strong market spillover with a moderate original market (see Fig. 7a, 7c, 7e, and 7g). However, the vertically integrated manufacturer with weaker customer perception will close component supply in a very large common market, i.e., in a near-monopoly original market with arbitrary spillover or in a relatively large original market with relatively strong spillover (see Fig. 7b, 7d, 7f, and 7h). Moreover, small original markets with high spillover and large original markets with low spillover encourage the vertically integrated manufacturer of weaker customer perception to open supply.

(2) In scenario $0 < \theta < 1$, improving the customer perception of $M_i$ or increasing the unit operation cost of $M_i$ or $M_e$ makes the vertically integrated manufacturer more likely to open component supply. It is because that the vertically integrated manufacturer becomes a "dual manufacturer" within the equilibrium open region (see Fig. 7c and Fig. 7e), and a smaller $\theta$ or a greater $m_e$ makes the increased profits in the product market offset the loss in the component market, thus expanding the equilibrium open region. Furthermore, when $m_i$ increases (see Fig. 7g), the equilibrium open region as a "dual manufacturer" expands (i.e., the region of $\hat{A} \geq \frac{16\theta(1-\theta)\pi_{M_e}^{0*}}{(\theta m_i - m_e)^2}$) and an extra equilibrium open region as a component manufacturer is built (i.e., the region of $arg\left\{\Pi_{M_i}^{0*} - \Pi_{M_{i+s,2}}(p_{i3}^{R_3\,*}, w_3^{R_3\,*}) = 0\right\} \leq \hat{A} < \frac{16\theta(1-\theta)\pi_{M_e}^{0*}}{(\theta m_i - m_e)^2}$).

(3) As for scenario $\theta > 1$, weaker customer perception of the vertically integrated manufacturer will narrow the equilibrium close region in moderate common markets and expand the equilibrium close region in large common markets (see Fig. 7d). The reason is that increasing $\theta$ improves the component wholesale price and the component demand of the vertically integrated manufacturer who becomes a component manufacturer in open-supply context, thus boosting its profits. When $\hat{A}$ is moderate (i.e., the region of $\hat{A} \leq \frac{4\theta^2 \pi_{M_e}^{0*}}{(\theta-1)(\theta^2 - m_e^2)}$), the increased profits lead to the curve $\Pi_{M_i}^{0*} = \Pi_{M_{i+s,2}}(p_{i4}^{R_6\,*}, w_4^{R_6\,*})$ shifts left, which reduces the equilibrium close region. When $\hat{A}$ is relatively large (i.e., the region of $\hat{A} > \frac{4\theta^2 \pi_{M_e}^{0*}}{(\theta-1)(\theta^2 - m_e^2)}$), increasing $\theta$ results a left shift of the curve $\hat{A} = \frac{4\theta^2 \pi_{M_e}^{0*}}{(\theta-1)(\theta^2 - m_e^2)}$, and the increased profits are still not enough to exceed the original profits except the region where the market spillover is high, thus the equilibrium close region expands.

(4) Counterintuitively, a greater operation cost advantage of the vertically integrated manufacturer narrows the equilibrium open region significantly (see Fig. 7f) while a greater operation cost disadvantage expands the equilibrium open region in moderate common markets (see Fig. 7h) in scenario $\theta > 1$. The explanation is that both the component wholesale price and the component demand decrease with $m_e$, which significantly lowers the profit of the vertically integrated manufacturer who only sells components and then reduces the equilibrium open region. However, increasing $m_i$ does not affect the profit of the vertically integrated manufacturer in open-supply context but lowers its original profit in close-supply context, which enlarges the equilibrium open region.

## 5. Conclusion

Motivated by a practical example in the new energy vehicle industry, we explore the competition behavior between a vertically integrated manufacturer and an end product manufacturer. For better development, the vertically integrated manufacturer may open the component supply and intensify the downstream competition with a product manufacturer who becomes its component partner. This paper investigates the strategic decision and pricing problem of two heterogeneous manufacturers with different customer perceived value and operation costs. To address the problem, we construct a three-stage game-theoretic model in which the vertically integrated manufacturer first announces its supply strategy, and sets the component price and product price followed by an exterior manufacturer making its sourcing and pricing decision. Our study provides the economic rationales for why BYD Auto announced the opening of battery supply and decided to supply batteries to manufacturers with lower customer perceived value. Moreover, our results reveal several main managerial insights, which are summarized as follows.

From the perspective of the vertically integrated manufacturer: (1) the vertically integrated manufacturer always closes the component supply within a small common market since it can not attract any exterior manufacturer after the implementation of open-supply strategy. (2) The vertically integrated manufacturer may open component supply to an exterior manufacturer who has a weaker customer perception and will close component supply when the exterior manufacturer has a stronger customer perception provided that the common market is sufficiently large. (3) The vertically integrated manufacturer may play one of the three roles (product manufacturer, component manufacturer, and "dual manufacturer") when possessing customer perception advantages. Whereas, if the exterior manufacturer has a customer perception advantage, then the vertically integrated manufacturer will shifts its role from a product manufacturer to a component manufacturer as the common market expands. Once an exterior manufacturer is attracted, the pricing decision of the vertically integrated manufacturer is always influenced by the operation cost and original profitability of its competitor, but is no matter with its own operation cost in Scenario $\theta > 1$. (4) The strategy maker must make a more accurate assessment of the market spillover when the original market share is moderate, because a high (low) spillover will result to open-supply (close-supply) strategy. (5) Compared with the scenario $0 < \theta < 1$, the request for the common market size is more relax in scenario $\theta > 1$ to attract an exterior manufacturer with the same operation cost and original profitability.

As to the insights for the exterior manufacturer, we find that: (1) if the common market is small, it is better not to change the component supplier. (2) The necessary condition for the exterior manufacturer to change a component supplier includes an upper bound on the component wholesale price and a lower bound on the end product sale price of its competitor. (3) When making price decisions in open-supply context, the exterior manufacturer does not need to consider the product price of the vertically integrated manufacturer who sets a relatively high product price, and can ignore the impact of the component wholesale price when the vertically integrated manufacturer sets a moderate product price. (4) The larger the common market size, the more likely the exterior manufacturer will change the component supplier in two scenarios, and the decrease of its operation cost will also increase the likelihood in scenario $\theta > 1$.

Further work on this topic can consider an investigation of two vertically integrated manufacturers' supply strategy decisions with heterogeneous consumers. If one of the vertically integrated manufacturers opens the component supply, does the other manufacturer use the open-supply component or the self-sufficient component? What role does the manufacturer play as the market changes when opening the supply? Moreover, if considering the relationship between the market spillover and brand building investment, what happens when the open-supply strategy implements? Hence, a rational vertically integrated manufacturer should trade off the investment cost and the investment income.